\documentclass[sigconf]{acmart}
\AtBeginDocument{%
  }

\copyrightyear{2026}
\acmYear{2026}
\setcopyright{cc}
\setcctype{by}
\acmConference[SIGIR '26]{Proceedings of the 49th International ACM SIGIR Conference on Research and Development in Information Retrieval}{July 20--24, 2026}{Melbourne, VIC, Australia}
\acmBooktitle{Proceedings of the 49th International ACM SIGIR Conference on Research and Development in Information Retrieval (SIGIR '26), July 20--24, 2026, Melbourne, VIC, Australia}
\acmDOI{10.1145/3805712.3808544}
\acmISBN{979-8-4007-2599-9/2026/07}

\usepackage{geometry}
\usepackage{hyperref}
\usepackage{enumitem}
\usepackage{booktabs}
\usepackage{graphicx}
\usepackage{xcolor}
\usepackage[normalem]{ulem}
\usepackage{amsmath}
\usepackage{tabularx}
\usepackage{array}
\usepackage{makecell}
\usepackage{multirow}
\usepackage{ragged2e}
\usepackage{tikz}
\usetikzlibrary{arrows.meta,positioning,calc}
\definecolor{revisionpurple}{RGB}{128,0,128}

\begin{document}

\title{LLM-Oriented Information Retrieval: A Denoising-First Perspective}

\author{Lu Dai}
\email{ldaiae@connect.ust.hk}
\affiliation{%
  \institution{Hong Kong University of Science and Technology}
  \city{Hong Kong SAR}
  \country{Hong Kong}
}
\author{Liang Sun}
\affiliation{%
  \institution{Hong Kong University of Science and Technology (Guangzhou)}
  \city{Guangzhou}
  \country{China}
}

\author{Fanpu Cao}
\affiliation{%
  \institution{Hong Kong University of Science and Technology (Guangzhou)}
  \city{Guangzhou}
  \country{China}
}

\author{Ziyang Rao}
\affiliation{%
  \institution{Hong Kong University of Science and Technology (Guangzhou)}
  \city{Guangzhou}
  \country{China}
}

\author{Cehao Yang}
\affiliation{%
  \institution{Hong Kong University of Science and Technology (Guangzhou)}
  \city{Guangzhou}
  \country{China}
}

\author{Hao Liu}
\email{liuh@ust.hk}
\affiliation{%
  \institution{Hong Kong University of Science and Technology (Guangzhou)}
  \city{Guangzhou}
  \country{China}
}

\author{Hui Xiong}
\email{xionghui@ust.hk}
\affiliation{%
  \institution{Hong Kong University of Science and Technology (Guangzhou)}
  \city{Guangzhou}
  \country{China}\\
  \institution{Hong Kong University of Science and Technology}
  \city{Hong Kong SAR}
  \country{Hong Kong}
}

\renewcommand\theadfont{\bfseries}

\begin{abstract}
Modern information retrieval (IR) is no longer consumed primarily by humans but increasingly by large language models (LLMs) via retrieval-augmented generation (RAG) and agentic search. 
Unlike human users, LLMs are constrained by limited attention budgets and are uniquely vulnerable to noise; misleading or irrelevant information is no longer just a nuisance, but a direct cause of hallucinations and reasoning failures. 
In this perspective paper, we argue that denoising—maximizing usable evidence density and verifiability within a context window—is becoming the primary bottleneck across the full information access pipeline. 
We conceptualize this paradigm shift through a four-stage framework of IR challenges: from \emph{inaccessible} to \emph{undiscoverable}, to \emph{misaligned}, and finally to \emph{unverifiable}. 
Furthermore, we provide a pipeline-organized taxonomy of signal-to-noise optimization techniques, spanning indexing, retrieval, context engineering, verification, and agentic workflow. We also present research works on information denoising in domains that rely heavily on retrieval such as lifelong assistant, coding agent, deep research, and multimodal understanding.
\end{abstract}




\maketitle
\renewcommand{\shortauthors}{Lu Dai et al.}

\section{Introduction}

\begin{figure*}[htb]
  \centering
  \includegraphics[width=0.95\linewidth]{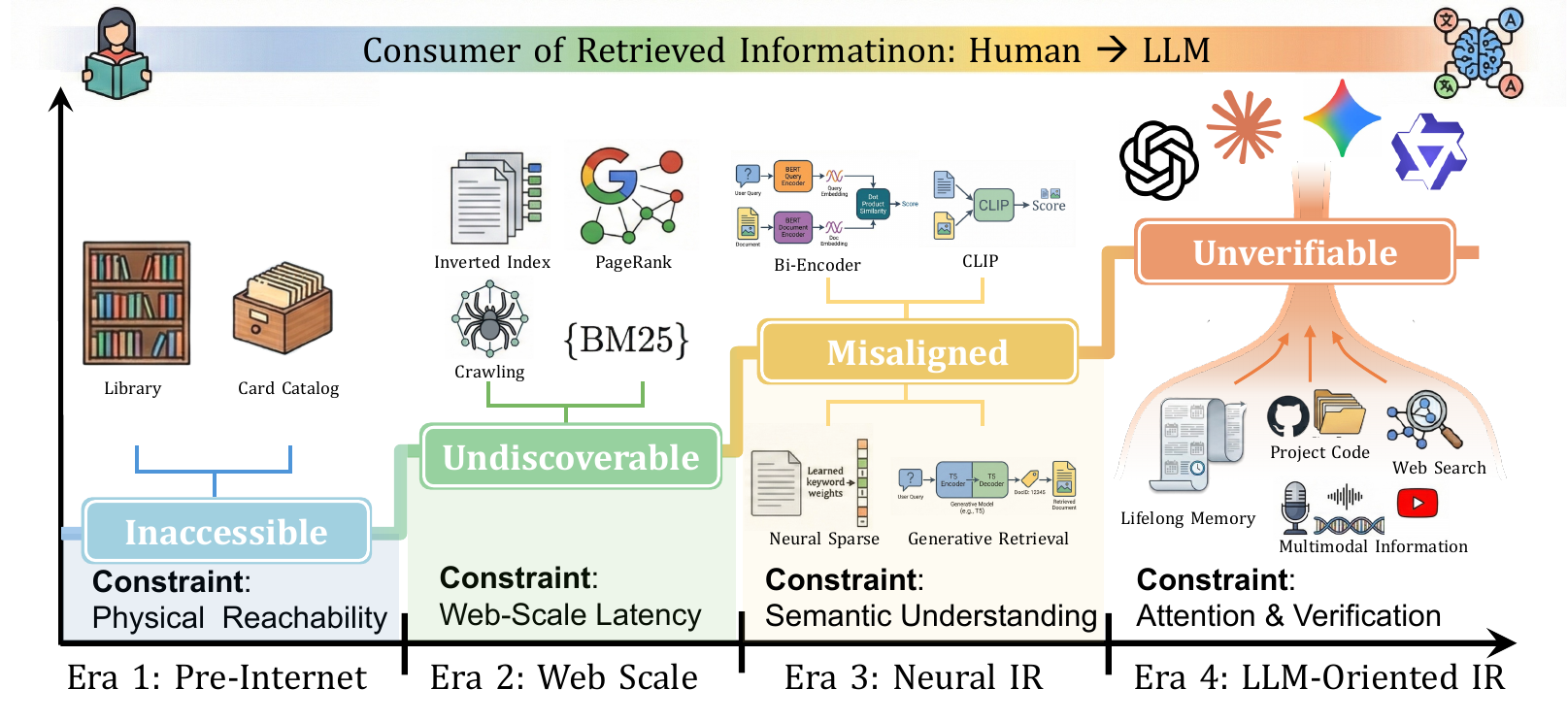}
  \caption{Challenge shifts in the history of IR.}
  \Description{A four-era timeline of information retrieval showing the dominant bottleneck shifting from inaccessible information in the pre-internet era, to undiscoverable information at web scale, to misaligned retrieval in neural IR, and finally to unverifiable evidence in LLM-oriented IR. Representative techniques and applications are placed along the timeline.}
  \label{fig:paradigm-shift}
\end{figure*}

Large‑language models (LLMs) have emerged as the new consumers of retrieved information and transformed information access. Rather than manually navigating web pages, users increasingly delegate their search, summarization and reasoning tasks to LLM agents \cite{zhai2025information, barbon2024large}. Accordingly, retrieval has become a crucial interleaving step in an LLM-centric task-solving pipeline, supplying the model with external knowledge to use in reasoning or generation. This paradigm – often referred to as retrieval-augmented generation (RAG) – tightly couples search with generation to improve factual coverage and answer quality \cite{lewis2020retrieval}.

As LLMs take on this role of information consumer, the objectives of information retrieval shift. Classic IR research prioritized metrics like recall and precision – focusing on minimizing missed relevant documents and down-ranking irrelevant ones. In LLM‑mediated systems, however, they are necessary but no longer sufficient.

One key bottleneck is verifiability. If the search component fails to find the correct supporting information or retrieves misleading information, even a powerful LLM cannot produce a correct and verifiable answer.  On the one hand, LLM-generated content is flooding the internet corpus itself. The proliferation of hallucinations makes attribution and trust harder than ever before. On the other hand, LLMs are sensitive to noise in context. Studies have found that misleading evidence in the context can be more detrimental to an LLM’s answer quality than missing evidence \cite{amiraz2025distracting}. These findings highlight that modern retrieval must focus not just on finding \textit{more} information, but on finding \textit{reliable} information that the LLM can faithfully use and verify as evidence.

Another critical challenge is the limited attention and context capacity of current LLMs. Even with the latest models featuring context windows extended to millions of tokens, LLMs struggle to fully utilize very large contexts, especially in reasoning tasks \cite{liu2024lost, chen2024benchmarking}. Empirical results confirm that beyond a certain point, adding more retrieved passages, especially irrelevant and redundant ones, yields diminishing or even negative returns. Furthermore, noisy or malicious content in the context can actively undermine the model’s reliability \cite{daiseper}. A well-known vulnerability is prompt injection, where adversarially crafted text in the retrieved context hijacks the model’s behavior or injects false information \cite{liu2023prompt}. This highlights that irrelevant or harmful information in the context window isn’t just a waste of computation resources – it can directly degrade output quality or even pose serious risks.

These observations suggest that the key challenge of modern retrieval is not to retrieve \textit{more} but to \textit{denoise}—to provide concise, high‑quality context that fits within the model’s attention budget. An LLM-based assistant might call a retriever frequently and aggregate many evidence snippets in one session; even a small fraction of noise can cascade through such a pipeline, leading to grounded hallucinations or degraded performance. While LLMs do exhibit some robustness, it is risky to rely on the model alone to filter out noise. Instead, the retrieval stage itself should take on the burden of ensuring a high signal-to-noise ratio in the provided context. 

In this paper, we present the perspective of denoising as the primary challenge for information retrieval in the era of LLM. 
We first characterize the challenge shift through the history of information retrieval into four stages and validate the importance of denoising with experimental evidence. 
We then present a comprehensive taxonomy of research efforts on denoising across the whole retrieval-augmented generation pipeline, as well as curating solutions in typical information-intensive and long-context application areas.
Finally, we summarize the solutions and limitations of current LLM-oriented information retrieval and propose potential future directions. 
Through this paper, we hope to shed light on the challenge shift of information retrieval in LLM era towards an emphasis on utility and verifiability, and inspire future innovations in this important research direction.

\section{Challenge shifts in the history of IR}
\label{sec:four-eras}

We conceptualize the evolution of IR as a progression of dominant bottlenecks—specific constraints that made information retrieval fail in distinct ways. As illustrated in Figure~\ref{fig:paradigm-shift}, the field has transitioned through four phases: from information \emph{inaccessible} to \emph{undiscoverable}, to \emph{misaligned}, and finally to \emph{unverifiable}.

\noindent\textbf{Era 1 (Pre-internet): Inaccessible under physical constraints}. \label{sec:era1} Before the World Wide Web, the ``evidence universe'' was inherently \emph{local}. Information was not primarily unranked, but \emph{unreachable}—constrained by geography, institutional access, and physical latency. In this era, the dominant bottleneck was availability rather than discoverability; the challenge was not selecting the best document, but physically acquiring any relevant document at all \cite{bush1945memex}.

\noindent\textbf{Era 2 (Web-scale IR): Undiscoverable under indexing scale}.
\label{sec:era2}
The web inverted the bottleneck: information became globally reachable but overwhelming. The challenge shifted to \emph{undiscoverability}—valid evidence existed but was buried under an ever-growing, unindexed corpus. 
In this stage, the core objective became efficient crawling, indexing, and ranking to surface signals from noise at interactive latencies.
Apart from relevance, quality estimation emerged as a critical filter, with PageRank leveraging hyperlink structures as a scalable proxy for authority to mitigate web noise~\cite{brin1998pagerank}.
Importantly, this era institutionalized large-scale evaluation infrastructure. Benchmarks and shared tasks (e.g., TREC) normalized reproducible experiments on large collections and anchored the field around measurable improvements under web-scale constraints.
Lexical matching (e.g., BM25) became the industrial standard, establishing a robust effectiveness–efficiency frontier that remains relevant today~\cite{robertson2009bm25}.

\noindent\textbf{Era 3 (Neural IR): Misaligned under semantic constraints}.
\label{sec:era3}
Once IR systems could reliably retrieve candidates at scale, the semantic barrier became the primary problem: 
even when evidence is retrieved, systems may not \emph{understand}, \emph{compose}, or \emph{reason with} it.
Besides, lexical overlaps often failed to capture user intent, leaving retrieved documents \emph{misaligned} with semantic needs. Research works in this era focused on bridging the semantic gap \cite{thakur2beir}.

Dense retrieval reduced the lexical mismatch between queries and relevant documents by learning continuous semantic representations (e.g., DPR) \cite{karpukhin2020dpr}.
Neural rankers improved relevance modeling but introduced steep inference costs; late-interaction architectures such as ColBERT offered a structural compromise between cross-encoder fidelity and bi-encoder efficiency \cite{khattab2020colbert}.
Meanwhile, neural sparse methods (e.g., SPLADE) reframed lexical matching itself as a learnable expansion mechanism, blurring the sparse-dense dichotomy \cite{formal2021splade}. 
Generative models also entered relevance modeling: MonoT5 cast reranking as a seq2seq prediction problem, using generative likelihood as a relevance signal \cite{nogueira2020monot5}.

In this era, IR systems began to deeply integrate with other NLP tasks under the retrieval-augmented generation scheme, setting the stage for end-to-end problem solving~\cite{izacard2021fid,lewis2020retrieval}.

\noindent\textbf{Era 4 (LLM-Oriented IR): Unverifiable context overload}.
\label{sec:era4_denoising_first}

We represent the current era not merely as an application shift, but as a fundamental redefinition of the retrieval objective driven by two simultaneous transformations.

First, LLMs have turned retrieval from a destination into an input channel. In retrieval-augmented generation, the consumer is no longer a human browsing for links, but a model requiring actionable evidence to drive generation~\cite{lewis2020retrieval, wang-etal-2024-searching}. Retrieval success is thus redefined by downstream utility: a document is only useful if it positively influences the model's reasoning trace~\cite{salemi2024erag, daiseper}.

Second, LLM-generated content is flooding the corpus. As the web becomes saturated with synthetic text, retrieval faces a crisis of epistemic integrity. The recursive training on generated content threatens model collapse ~\cite{shumailov2024modelcollapse}, creating a homogeneous and potentially adversarial search space where source bias and noise actively compete with factual signals~\cite{dai-etal-2024-cocktail}.

These shifts expose a fundamental friction: while traditional IR strives for high recall, LLMs struggle to filter the resulting noise from their limited attention span. We identify three intrinsic vulnerabilities where retrieval artifacts directly degrade reasoning:

\noindent\textit{Observation 1: Fragmentation brings in conflicts.} Retrieving and assembling snippets from disparate sources often strips away their original discourse context, such as temporal validity or conditional scope. This de-contextualization creates a "Frankenstein" context where conflicting or outdated information appears semantically relevant. Such semantic noise actively competes with correct evidence for attention, confusing the model's selection mechanism even when the context length is manageable~\cite{wang2025astute, yoran2024making}.

\noindent\textit{Observation 2: Context Dilution}. Extended context windows do not guarantee utilization. As the retrieved context grows, useful information becomes buried or "lost-in-the-middle," exhausting the model's attention budget. This dilution effect often leads to reasoning failures even when the answer is present in the context~\cite{liu2024lost, bai2024longbench}.

\noindent\textit{Observation 3: Cascading Failures}. With the rise of agentic and multi-step reasoning workflows, the cost of retrieval errors has escalated. A single misleading or conflicting snippet can propagate through the reasoning chain, converting retrieval noise into compounded hallucinations that undermine the entire task execution~\cite{niu-etal-2024-ragtruth}.

Consequently, we argue that the primary bottleneck in Era 4 has shifted from \emph{access} to \emph{utility}. To prevent these upstream noises from disrupting downstream reasoning, the retrieval system must evolve into a \emph{noise gate}, prioritizing the maximization of signal-to-noise ratio and verifiability over raw recall.
\begin{figure}[b]
  \centering
  \includegraphics[width=\linewidth]{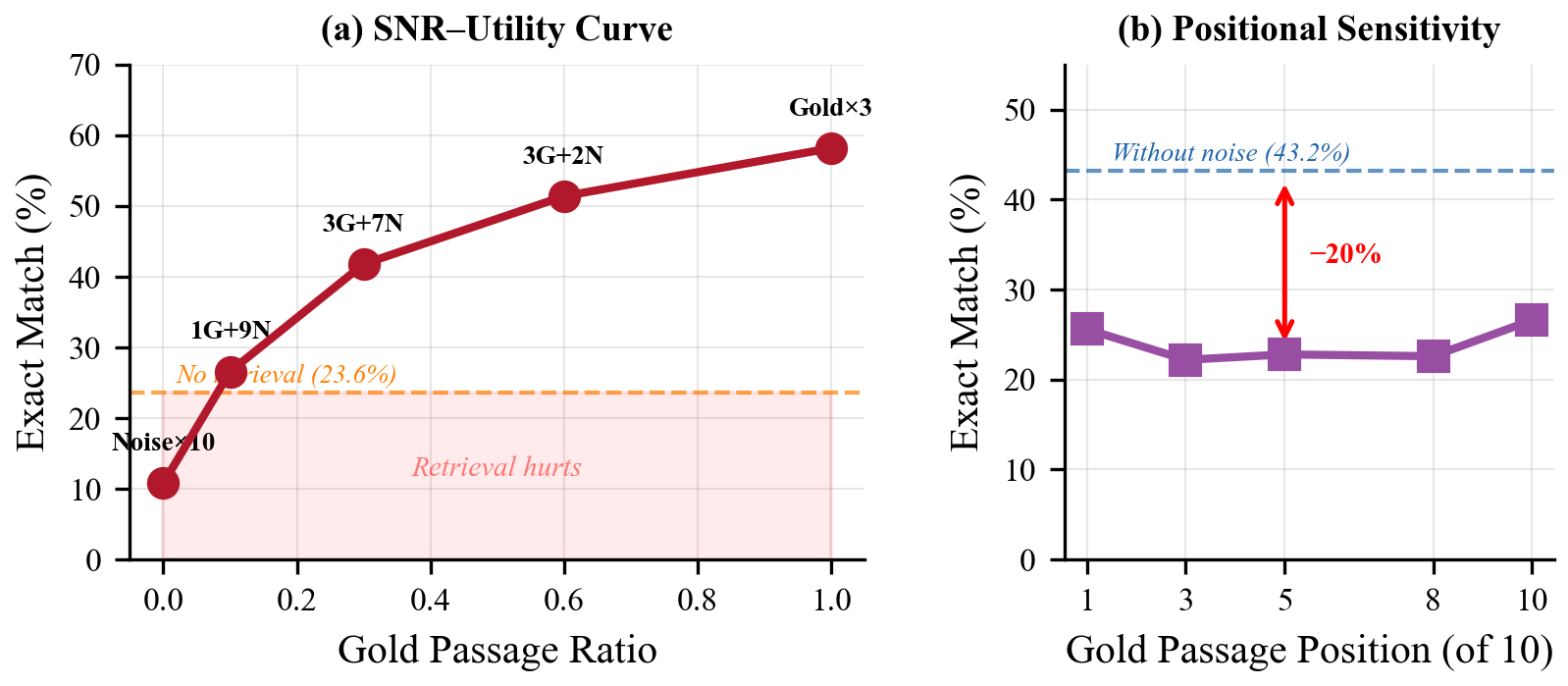}
  \caption{Empirical validation of the denoising-first perspective on NQ. \textbf{(a)} SNR--utility curve: as the gold passage ratio decreases, EM degrades sharply. \textbf{(b)} Positional sensitivity: the $\sim$20\% drop from noise $>>$ the $\sim$4\% positional variation.}
  \Description{A two-panel empirical figure on Natural Questions. The left panel plots exact-match accuracy against gold passage ratio and shows a steep monotonic drop as noise increases. The right panel compares gold-passage positions inside a context with one gold and nine noise passages, showing only a small positional effect relative to the much larger degradation caused by noise.}
  \label{fig:denoising-analysis}
\end{figure}

\subsection{Empirical Validation: Noise as Bottleneck}
\label{sec:empirical_validation}

We validated the "denoising-first" perspective using LLaMA-2-7B-Chat on 500 Natural Questions (NQ) samples, each paired with 100 DPR-retrieved passages (labeled as gold or noise).

\noindent\textbf{Impact of Context SNR.}
We systematically vary the signal-to-noise ratio of retrieved context. 
As shown in Figure~\ref{fig:denoising-analysis}(a), EM improves monotonically from 47.0\% to 61.0\% when increasing gold passages from 1 to 5. 
However, noise erodes this gain rapidly: holding gold passages fixed at 3, adding 2 and 7 noise passages reduces EM to 51.4\% and 41.8\%, respectively. 
When a single gold passage is buried among 9 noise passages (SNR$=$0.10), EM falls to 26.6\%, barely above the closed-book baseline of 23.6\%. 
Pure noise contexts yield only 8.0\% EM—far \emph{worse} than no retrieval—demonstrating that noise actively impairs the model's parametric memory.

\noindent\textbf{Positional Bias vs. Noise.}
Fixing the context to 1 gold and 9 noise passages while varying the gold passage's position (Figure~\ref{fig:denoising-analysis}(b)), we find that noise-induced degradation dominates positional effects. 
A single gold passage alone achieves 43.2\% EM, but adding 9 noise passages collapses performance to 22.2--26.6\% regardless of position. A mild U-shaped positional pattern is visible but dwarfed by the noise effect.

\section{Denoising in LLM-Oriented Retrieval: A Method Taxonomy}
\label{sec:denoising_taxonomy}

In the context of LLM-oriented IR, the retrieval objective fundamentally shifts from prioritizing recall to acting as a \textbf{noise gate}. 
We use \emph{noise} to refer to any retrieved content that consumes budget (tokens, latency, attention) without helping---or that actively misleads the model. 
Drawing on the shift described in Section \ref{sec:four-eras}, we categorize the primary entry points of noise in LLM-oriented IR into three distinct classes:

\noindent\textbf{Corpus-level noise.}
Noise originates upstream within the knowledge source itself, where indices are increasingly contaminated by duplicates, template spam, or outdated information \cite{liang2025widespread}. 
Recent studies highlight a growing risk of ``spiral'' feedback loops, where AI-generated content becomes overrepresented in retrieval pools, potentially destabilizing provenance \cite{chen2024spiral}. 
Furthermore, even minor injections of adversarial misinformation or low-credibility evidence can severely degrade open-domain QA, as standard retrievers lack the mechanisms to verify truthfulness before indexing \cite{pan2023attacking}.

\noindent\textbf{Retriever-level noise.}
Even with a clean corpus, retrievers introduce noise by surfacing ``hard distractors''—passages that are topically relevant but factually non-supporting or misleading \cite{yoran2024making}. 
Dense retrievers are particularly prone to these semantic ``near misses,'' which occupy the embedding space near the query but fail to answer it \cite{amiraz2025distracting,dai2024improve}. 
Research indicates that these non-informative contexts actively divert the model's attention, leading to distraction-induced hallucinations and reduced accuracy even when correct evidence is present \cite{liu2024lost}.

\noindent\textbf{Context construction noise.} Noise is often amplified in the last mile between retrieval and generation, when candidates are chunked, merged, and ordered into a single prompt. First, structural and positional noise stems from the mismatch between retrieval ranking and LLM attention; naive concatenation often buries critical evidence in the ``lost-in-the-middle'' zone, rendering it effectively invisible to the model \cite{liu2024lost,wang2025documentsegmentation}. 
Second, evidence incoherence arises when the retriever aggregates conflicting or redundant passages; this incoherence increases parametric uncertainty and triggers hallucinations, requiring diversity-aware compression rather than simple concatenation \cite{xie2024adaptive,wang2025conflicting}. 
Third, the retrieval channel introduces adversarial noise via indirect prompt injections, where malicious instructions embedded in untrusted documents hijack the generation logic, breaching the system's safety alignment \cite{greshake2023not}. 
Consequently, the IR system must treat evidence assembly as an active denoising stage—responsible for curating, ordering, and sanitizing the prompt—rather than a passive data hand-off.

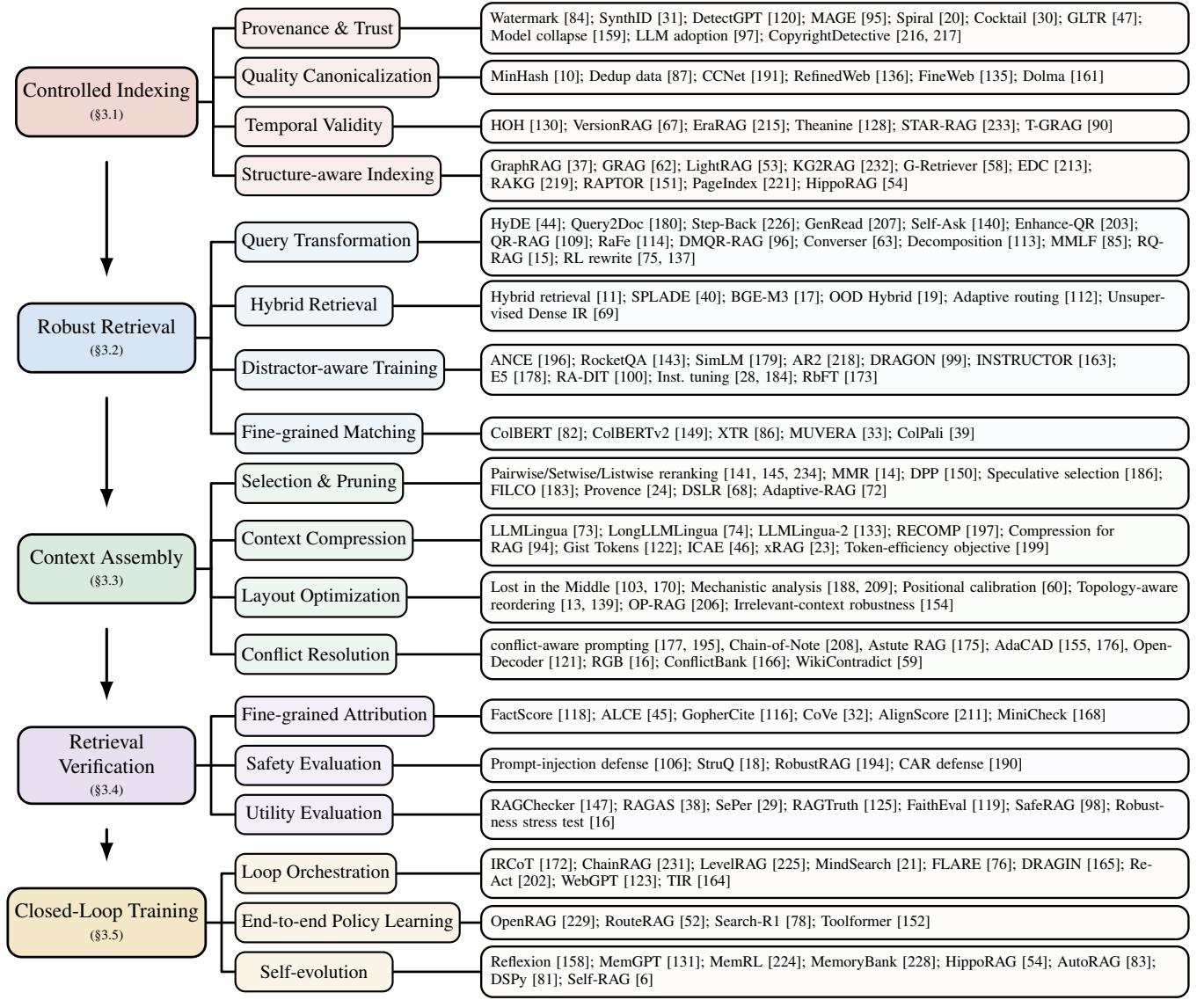
\begin{figure*}[t]
\centering
\resizebox{\textwidth}{!}{%
\begin{tikzpicture}[x=1cm,y=1cm,>=Latex,line cap=round,line join=round]

\definecolor{s1}{HTML}{EFD7D2}
\definecolor{s1l}{HTML}{F9EEEB}
\definecolor{s2}{HTML}{D7E6F4}
\definecolor{s2l}{HTML}{EEF5FB}
\definecolor{s3}{HTML}{D9EBDD}
\definecolor{s3l}{HTML}{EEF6F0}
\definecolor{s4}{HTML}{E8DFF1}
\definecolor{s4l}{HTML}{F3EEF8}
\definecolor{s5}{HTML}{F5E8C9}
\definecolor{s5l}{HTML}{FBF5E8}

\newcommand{\stageW}{1.8cm}
\newcommand{\stageH}{0.7cm}       
\newcommand{\conceptW}{1.6cm}
\newcommand{\conceptH}{0.4cm}     
\newcommand{\leafW}{7.0cm}        
\newcommand{\colStage}{0.0}       
\newcommand{\colConcept}{1.3}     
\newcommand{\colLeafL}{3.8}       
\newcommand{\secref}{\fontsize{4}{5}\selectfont}

\newcommand{\stageGapOneTwo}{2.4}     
\newcommand{\stageGapTwoThree}{2.35}   
\newcommand{\stageGapThreeFour}{2.0}  
\newcommand{\stageGapFourFive}{1.6}   

\pgfmathsetmacro{\stageYone}{0}
\pgfmathsetmacro{\stageYtwo}{\stageYone-\stageGapOneTwo}
\pgfmathsetmacro{\stageYthree}{\stageYtwo-\stageGapTwoThree}
\pgfmathsetmacro{\stageYfour}{\stageYthree-\stageGapThreeFour}
\pgfmathsetmacro{\stageYfive}{\stageYfour-\stageGapFourFive}

\newcommand{\csepOne}{0.50}    
\newcommand{\csepTwo}{0.65}    
\newcommand{\csepThree}{0.58}  
\newcommand{\csepFour}{0.50}   
\newcommand{\csepFive}{0.50}   

\tikzset{
  flow/.style={-{Latex[length=2.0mm,width=1.4mm]},line width=0.8pt,draw=black},
  edge/.style={line width=0.7pt,draw=black},
  stage/.style={
    draw=black,rounded corners=1.4mm,line width=0.7pt,
    minimum width=\stageW,minimum height=\stageH,align=center,
    inner sep=2pt,
    font=\fontfamily{ptm}\selectfont\fontsize{6}{7.2}\selectfont,text=black
  },
  concept/.style={
    draw=black,rounded corners=1.0mm,line width=0.55pt,
    minimum width=\conceptW,minimum height=\conceptH,align=left,anchor=west,
    inner sep=1.8pt,
    font=\fontfamily{ptm}\selectfont\fontsize{5.5}{6.5}\selectfont,text=black
  },
  leaf/.style={
    draw=black,rounded corners=1.2mm,line width=0.6pt,
    text width=\leafW,align=left,inner sep=2.8pt,anchor=west,
    font=\fontfamily{ptm}\selectfont\fontsize{4.5}{4.5}\selectfont,text=black
  },
  stage1/.style={stage,fill=s1},
  stage2/.style={stage,fill=s2},
  stage3/.style={stage,fill=s3},
  stage4/.style={stage,fill=s4},
  stage5/.style={stage,fill=s5},
  concept1/.style={concept,fill=s1l},
  concept2/.style={concept,fill=s2l},
  concept3/.style={concept,fill=s3l},
  concept4/.style={concept,fill=s4l},
  concept5/.style={concept,fill=s5l},
  leaf1/.style={leaf,fill=s1l!25!white},
  leaf2/.style={leaf,fill=s2l!25!white},
  leaf3/.style={leaf,fill=s3l!25!white},
  leaf4/.style={leaf,fill=s4l!25!white},
  leaf5/.style={leaf,fill=s5l!25!white}
}

\node[stage1] (s1) at (\colStage, \stageYone)
  {Controlled Indexing\\[-1pt]\secref(\S\ref{sec:controlled_indexing})};
\node[stage2] (s2) at (\colStage, \stageYtwo)
  {Robust Retrieval\\[-1pt]\secref(\S\ref{sec:robust_retrieval})};
\node[stage3] (s3) at (\colStage, \stageYthree)
  {Context Assembly\\[-1pt]\secref(\S\ref{sec:context_assembly})};
\node[stage4] (s4) at (\colStage, \stageYfour)
  {Retrieval\\Verification\\[-1pt]\secref(\S\ref{sec:faithfulness_verification})};
\node[stage5] (s5) at (\colStage, \stageYfive)
  {Closed-Loop Training\\[-1pt]\secref(\S\ref{sec:agentic_loops})};

\draw[flow] ($(s1.south)+(0,-0.25)$) -- ($(s2.north)+(0,0.25)$);
\draw[flow] ($(s2.south)+(0,-0.25)$) -- ($(s3.north)+(0,0.25)$);
\draw[flow] ($(s3.south)+(0,-0.25)$) -- ($(s4.north)+(0,0.25)$);
\draw[flow] ($(s4.south)+(0,-0.25)$) -- ($(s5.north)+(0,0.25)$);

\pgfmathsetmacro{\soney}{\stageYone} 

\node[concept1] (s1c1) at (\colConcept, \soney+1.5*\csepOne) {Provenance \& Trust};
\node[concept1] (s1c2) at (\colConcept, \soney+0.5*\csepOne) {Quality Canonicalization};
\node[concept1] (s1c3) at (\colConcept, \soney-0.5*\csepOne) {Temporal Validity};
\node[concept1] (s1c4) at (\colConcept, \soney-1.5*\csepOne) {Structure-aware Indexing};

\coordinate (s1h) at ($(s1.east)+(0.15,0)$);
\draw[edge] (s1.east) -- (s1h);
\draw[edge] (s1h) |- (s1c1.west);
\draw[edge] (s1h) |- (s1c2.west);
\draw[edge] (s1h) |- (s1c3.west);
\draw[edge] (s1h) |- (s1c4.west);

\node[leaf1] (s1l1) at (\colLeafL, \soney+1.5*\csepOne)
  {Watermark~\cite{kirchenbauer2023watermark}; SynthID~\cite{dathathri2024synthid}; DetectGPT~\cite{mitchell2023detectgpt}; MAGE~\cite{li-etal-2024-mage}; Spiral~\cite{chen2024spiral}; Cocktail~\cite{dai-etal-2024-cocktail}; GLTR~\cite{gehrmann-etal-2019-gltr}; Model collapse~\cite{shumailov2024modelcollapse}; LLM adoption~\cite{liang2025widespread}; CopyrightDetective~\cite{zhang2026copyright, zhang-etal-2025-isacl}};
\node[leaf1] (s1l2) at (\colLeafL, \soney+0.5*\csepOne)
  {MinHash~\cite{broder1997resemblance}; Dedup data~\cite{lee-etal-2022-deduplicating}; CCNet~\cite{wenzek-etal-2020-ccnet}; RefinedWeb~\cite{penedo2023refinedweb}; FineWeb~\cite{penedo2024fineweb}; Dolma~\cite{soldaini2024dolma}};
\node[leaf1] (s1l3) at (\colLeafL, \soney-0.5*\csepOne)
  {HOH~\cite{ouyang-etal-2025-hoh}; VersionRAG~\cite{huwiler2025versionrag}; EraRAG~\cite{zhang2025erarag}; Theanine~\cite{ong2025theanine}; STAR-RAG~\cite{zhu2025right}; T-GRAG~\cite{li2025tgrag}};
\node[leaf1] (s1l4) at (\colLeafL, \soney-1.5*\csepOne)
  {GraphRAG~\cite{edge2024localtoglobal}; GRAG~\cite{hu2024grag}; LightRAG~\cite{guo2024lightrag}; KG2RAG~\cite{zhu-etal-2025-knowledge}; G-Retriever~\cite{he2024g-retriever}; EDC~\cite{zhang-soh-2024-extract}; RAKG~\cite{zhang2025rakg}; RAPTOR~\cite{sarthi2024raptor}; PageIndex~\cite{vectifyai2025pageindex}; HippoRAG~\cite{hipporag}};

\draw[edge] (s1c1.east) -- (s1l1.west);
\draw[edge] (s1c2.east) -- (s1l2.west);
\draw[edge] (s1c3.east) -- (s1l3.west);
\draw[edge] (s1c4.east) -- (s1l4.west);

\pgfmathsetmacro{\stwoy}{\stageYtwo} 

\node[concept2] (s2c1) at (\colConcept, \stwoy+1.5*\csepTwo) {Query Transformation};
\node[concept2] (s2c2) at (\colConcept, \stwoy+0.5*\csepTwo) {Hybrid Retrieval};
\node[concept2] (s2c3) at (\colConcept, \stwoy-0.5*\csepTwo) {Distractor-aware Training};
\node[concept2] (s2c4) at (\colConcept, \stwoy-1.5*\csepTwo) {Fine-grained Matching};

\coordinate (s2h) at ($(s2.east)+(0.15,0)$);
\draw[edge] (s2.east) -- (s2h);
\draw[edge] (s2h) |- (s2c1.west);
\draw[edge] (s2h) |- (s2c2.west);
\draw[edge] (s2h) |- (s2c3.west);
\draw[edge] (s2h) |- (s2c4.west);

\node[leaf2] (s2l1) at (\colLeafL, \stwoy+1.5*\csepTwo)
  {HyDE~\cite{gao2023precise}; Query2Doc~\cite{wang2023querydoc}; Step-Back~\cite{zheng2024stepback}; GenRead~\cite{yu2023generate}; Self-Ask~\cite{press2023selfask}; Enhance-QR~\cite{ye2023enhancing}; QR-RAG~\cite{ma-etal-2023-query}; RaFe~\cite{mao-etal-2024-rafe}; DMQR-RAG~\cite{li2024dmqr}; Converser~\cite{huang2023converser}; Decomposition~\cite{mao2023large}; MMLF~\cite{kuo-etal-2025-mmlf}; RQ-RAG~\cite{chan2024rqrag}; RL rewrite~\cite{peng2024large,jiang2025deepretrieval}};
\node[leaf2] (s2l2) at (\colLeafL, \stwoy+0.5*\csepTwo)
  {Hybrid retrieval~\cite{bruch2023analysis}; SPLADE~\cite{formal2021splade}; BGE-M3~\cite{chen2024bgem3}; OOD Hybrid~\cite{chen2022outofdomain}; Adaptive routing~\cite{mallen-etal-2023-whennot}; Unsupervised Dense IR~\cite{izacard2021unsupervised}};
\node[leaf2] (s2l3) at (\colLeafL, \stwoy-0.5*\csepTwo)
  {ANCE~\cite{xiong2020ance}; RocketQA~\cite{qu2021rocketqa}; SimLM~\cite{wang-etal-2023-simlm}; AR2~\cite{zhang2022ar2}; DRAGON~\cite{lin2023dragon}; INSTRUCTOR~\cite{su-etal-2023-instructor}; E5~\cite{wang2022e5}; RA-DIT~\cite{lin2024radit}; Inst.\ tuning~\cite{wang2025foundation,dai2024improve}; RbFT~\cite{tu2025rbft}};
\node[leaf2] (s2l4) at (\colLeafL, \stwoy-1.5*\csepTwo)
  {ColBERT~\cite{khattab2020colbert}; ColBERTv2~\cite{santhanam2021colbertv2}; XTR~\cite{lee2023xtr}; MUVERA~\cite{dhulipala2024muvera}; ColPali~\cite{faysse2025colpali}};

\draw[edge] (s2c1.east) -- (s2l1.west);
\draw[edge] (s2c2.east) -- (s2l2.west);
\draw[edge] (s2c3.east) -- (s2l3.west);
\draw[edge] (s2c4.east) -- (s2l4.west);

\pgfmathsetmacro{\sthreey}{\stageYthree} 

\node[concept3] (s3c1) at (\colConcept, \sthreey+1.5*\csepThree) {Selection \& Pruning};
\node[concept3] (s3c2) at (\colConcept, \sthreey+0.5*\csepThree) {Context Compression};
\node[concept3] (s3c3) at (\colConcept, \sthreey-0.5*\csepThree) {Layout Optimization};
\node[concept3] (s3c4) at (\colConcept, \sthreey-1.5*\csepThree) {Conflict Resolution};

\coordinate (s3h) at ($(s3.east)+(0.15,0)$);
\draw[edge] (s3.east) -- (s3h);
\draw[edge] (s3h) |- (s3c1.west);
\draw[edge] (s3h) |- (s3c2.west);
\draw[edge] (s3h) |- (s3c3.west);
\draw[edge] (s3h) |- (s3c4.west);

\node[leaf3] (s3l1) at (\colLeafL, \sthreey+1.5*\csepThree)
  {Pairwise/Setwise/Listwise reranking~\cite{qin2024pairwise,zhuang2024setwise,reddy2024first};
   MMR~\cite{DBLP:conf/sigir/CarbonellG98};
   DPP~\cite{sarlak2026reliabilityaware};
   Speculative selection~\cite{wang2025speculative};
   FILCO~\cite{wang2023filco};
   Provence~\cite{chirkova2025provence};
   DSLR~\cite{hwang2024dslr};
   Adaptive-RAG~\cite{jeong2024adaptiverag}};
\node[leaf3] (s3l2) at (\colLeafL, \sthreey+0.5*\csepThree)
  {LLMLingua~\cite{jiang-etal-2023-llmlingua}; LongLLMLingua~\cite{jiang-etal-2024-longllmlingua}; LLMLingua-2~\cite{pan2024llmlingua2}; RECOMP~\cite{xu2024recomp}; Compression for RAG~\cite{li-etal-2023-compressing}; Gist Tokens~\cite{mu2023learning}; ICAE~\cite{ge2024incontext}; xRAG~\cite{cheng2024xrag}; Token-efficiency objective~\cite{DBLP:conf/iclr/XuPSC25}};
\node[leaf3] (s3l3) at (\colLeafL, \sthreey-0.5*\csepThree)
  {Lost in the Middle~\cite{liu2024lost,tian2025distance};
   Mechanistic analysis~\cite{yu2024mitigate,wang2025pine};
   Positional calibration~\cite{hsieh2024found};
   Topology-aware reordering~\cite{peysakhovich2023attention,byerly2025gold};
   OP-RAG~\cite{yu2024defense};
   Irrelevant-context robustness~\cite{shi2023large}};
\node[leaf3] (s3l4) at (\colLeafL, \sthreey-1.5*\csepThree)
  {conflict-aware prompting~\cite{xie2024adaptive,wang2025conflicting},
   Chain-of-Note~\cite{yu2023chainofnote},
   Astute RAG~\cite{wang2025astute};
   AdaCAD~\cite{shi2024trusting,wang2025adacad},
   OpenDecoder~\cite{mo2026opendecoder};
   RGB~\cite{chen2024benchmarking};
   ConflictBank~\cite{su2024conflictbank};
   WikiContradict~\cite{hou2024wikicontradict}};
\draw[edge] (s3c1.east) -- (s3l1.west);
\draw[edge] (s3c2.east) -- (s3l2.west);
\draw[edge] (s3c3.east) -- (s3l3.west);
\draw[edge] (s3c4.east) -- (s3l4.west);

\pgfmathsetmacro{\sfoury}{\stageYfour} 

\node[concept4] (s4c1) at (\colConcept, \sfoury+\csepFour)  {Fine-grained Attribution};
\node[concept4] (s4c2) at (\colConcept, \sfoury)           {Safety Evaluation};
\node[concept4] (s4c3) at (\colConcept, \sfoury-\csepFour)  {Utility Evaluation};

\coordinate (s4h) at ($(s4.east)+(0.15,0)$);
\draw[edge] (s4.east) -- (s4h);
\draw[edge] (s4h) |- (s4c1.west);
\draw[edge] (s4h) |- (s4c2.west);
\draw[edge] (s4h) |- (s4c3.west);

\node[leaf4] (s4l1) at (\colLeafL, \sfoury+\csepFour)
  {FactScore~\cite{min2023factscore}; ALCE~\cite{gao-etal-2023-enabling}; GopherCite~\cite{menick2022gophercite}; CoVe~\cite{dhuliawala2024chainofverification}; AlignScore~\cite{zha2023alignscore}; MiniCheck~\cite{tang2024minicheck}};
\node[leaf4] (s4l2) at (\colLeafL, \sfoury)
  {Prompt-injection defense~\cite{liu2024promptinjection}; StruQ~\cite{chen2025struq}; RobustRAG~\cite{xiang2025robustrag}; CAR defense~\cite{weller2024defending}};
\node[leaf4] (s4l3) at (\colLeafL, \sfoury-\csepFour)
  {RAGChecker~\cite{ru2024ragchecker}; RAGAS~\cite{es2023ragas}; SePer~\cite{daiseper}; RAGTruth~\cite{niu-etal-2024-ragtruth}; FaithEval~\cite{mingfaitheval}; SafeRAG~\cite{liang2025saferag}; Robustness stress test~\cite{chen2024benchmarking}};

\draw[edge] (s4c1.east) -- (s4l1.west);
\draw[edge] (s4c2.east) -- (s4l2.west);
\draw[edge] (s4c3.east) -- (s4l3.west);

\pgfmathsetmacro{\sfivey}{\stageYfive} 

\node[concept5] (s5c1) at (\colConcept, \sfivey+\csepFive) {Loop Orchestration};
\node[concept5] (s5c3) at (\colConcept, \sfivey) {End-to-end Policy Learning};
\node[concept5] (s5c4) at (\colConcept, \sfivey-\csepFive) {Self-evolution};

\coordinate (s5h) at ($(s5.east)+(0.15,0)$);
\draw[edge] (s5.east) -- (s5h);
\draw[edge] (s5h) |- (s5c1.west);
\draw[edge] (s5h) |- (s5c3.west);
\draw[edge] (s5h) |- (s5c4.west);

\node[leaf5] (s5l1) at (\colLeafL, \sfivey+\csepFive)
  {IRCoT~\cite{trivedi2023ircot}; ChainRAG~\cite{zhu2025chainrag}; LevelRAG~\cite{zhang2025levelrag}; MindSearch~\cite{chen2024mindsearch}; FLARE~\cite{jiang2023active}; DRAGIN~\cite{su2024dragin}; ReAct~\cite{yao2022react}; WebGPT~\cite{nakano2021webgpt}; TIR~\cite{su2026beyond}};
\node[leaf5] (s5l3) at (\colLeafL, \sfivey)
  {OpenRAG~\cite{zhou2025openrag}; RouteRAG~\cite{guo2025routerag}; Search-R1~\cite{searchr1}; Toolformer~\cite{schick2023toolformer}};
\node[leaf5] (s5l4) at (\colLeafL, \sfivey-\csepFive)
  {Reflexion~\cite{shinn2023reflexion}; MemGPT~\cite{packer2023memgpt}; MemRL~\cite{zhang2026memrl}; MemoryBank~\cite{memorybank}; HippoRAG~\cite{hipporag}; AutoRAG~\cite{kim2024autorag}; DSPy~\cite{khattab2024dspy}; Self-RAG~\cite{asai2024selfrag}};

\draw[edge] (s5c1.east) -- (s5l1.west);
\draw[edge] (s5c3.east) -- (s5l3.west);
\draw[edge] (s5c4.east) -- (s5l4.west);

\end{tikzpicture}%
}
\caption{A multi-level denoising taxonomy aligned with the five-stage Section~3 pipeline:
Controlled Indexing (\S\ref{sec:controlled_indexing}),
Robust Retrieval (\S\ref{sec:robust_retrieval}),
Context Assembly (\S\ref{sec:context_assembly}),
Retrieval Verification (\S\ref{sec:faithfulness_verification}),
and Closed-Loop Training (\S\ref{sec:agentic_loops}).}
\Description{A large stage-aligned taxonomy diagram for denoising in LLM-oriented retrieval. It organizes methods into five sequential stages: controlled indexing, robust retrieval, context assembly, faithfulness verification, and closed-loop training. Each stage contains several concepts connected to example methods and representative papers.}
\label{fig:denoising-taxonomy}
\end{figure*}

To systematically address these failure modes, we organize our taxonomy of denoising methods aligning with the lifecycle of information flow. 
The following sections review interventions at the level of \textbf{Representation} (purifying the index), \textbf{Search} (filtering via precision ranking), \textbf{Context Assembly} (optimizing input composition), and \textbf{Verification} (auditing the generative output). Finally, we examine \textbf{Agentic Loops}, where feedback dynamically refines the signal through iterative retrieval and self-correction.
Figure~\ref{fig:denoising-taxonomy} summarizes this taxonomy in a stage-aligned pipeline view.

\subsection{Controlled Indexing}
\label{sec:controlled_indexing}
This subsection surveys controlled indexing---index-time interventions that reduce noise by
(i) regulating what enters the retrievable universe and
(ii) attaching \emph{conditionable} representations and metadata (e.g., trust, time, structure, safety) that downstream retrieval and context assembly can reliably exploit.
Governed indexing sets the retrieval quality ceiling, as downstream filtering cannot fully eliminate a polluted index's residual noise and overhead.

\noindent\textbf{(1) Provenance and Trust Stratification.}
Rather than treating the corpus as a monolithic pool, effective denoising begins by stratifying the index based on verifyable provenance.
This involves capturing granular metadata—such as publisher authority, extraction timestamps, and cryptographic signatures (e.g., C2PA standards \cite{c2pa2025spec})—to enable \emph{trust-conditioned retrieval}, where the search space can be dynamically restricted to high-credibility strata.
A critical challenge in the LLM era is distinguishing human insights from machine-generated content to prevent degradation caused by widespread AI content on the open web \cite{liang2025widespread,chen2024spiral,dai-etal-2024-cocktail, shumailov2024modelcollapse}.
To this end, indexing pipelines must integrate \emph{synthetic attribution} mechanisms: strictly cataloging proactive watermarks \cite{kirchenbauer2023watermark,dathathri2024synthid}; embedding post-hoc detection scores \cite{mitchell2023detectgpt,li-etal-2024-mage,gehrmann-etal-2019-gltr} as filterable metadata; and flagging copyrighted or otherwise high-risk content for downstream gating \cite{zhang2026copyright,zhang-etal-2025-isacl}.
By operationalizing provenance as an admission criterion rather than an afterthought, the system prevents low-credibility signals from ever competing for the retriever's attention.

\noindent\textbf{(2) Quality Filtering and Canonicalization.} Data filtering is not only important for LLM pretraining, but also for retrieval index building, where even small fractions of irrelevant or low-quality passages have been shown to substantially degrade RAG generation quality \cite{cuconasu2024power}.
To maximize the \emph{information density} of the index, this family of methods eliminates semantic redundancy and low-quality artifacts that otherwise act as distractors in the embedding space.
Deduplication is one of the levers: surface-form resemblance measures (e.g., MinHash) prune near-duplicates \cite{broder1997resemblance,lee-etal-2022-deduplicating}, while embedding-based semantic deduplication (e.g., SemDeDup) further collapses paraphrastic clusters that lexical hashing misses \cite{abbas2024semdedup}, ensuring the index stores only canonical prototypes rather than repetitive template variations and preventing budget waste during context assembly.
Furthermore, data sanitation pipelines established for LLM pre-training provide transferable primitives for index purification.
These techniques span language identification, heuristic filtering, and safety scoring as seen in works like RefinedWeb~\cite{wenzek-etal-2020-ccnet, penedo2023refinedweb,penedo2024fineweb, soldaini2024dolma, li2024datacomp}.
Adapting these protocols ensures that the retrieval pool is chemically pure, reducing the probability of hallucination-inducing nonsense.

\noindent\textbf{(3) Temporal Validity Management.}
In dynamic environments, semantic relevance no longer implies factual currency, motivating indexing pipelines that treat \emph{temporal validity} as a first-class admission constraint rather than a static snapshot \cite{ouyang-etal-2025-hoh}.
Recent designs realize this through time-aware graph structures: VersionRAG and EraRAG maintain explicit version graphs with incremental updates that avoid costly full rebuilds \cite{huwiler2025versionrag,zhang2025erarag}, while STAR-RAG summarizes graph snapshots and T-GRAG attaches temporal edges to disambiguate overlapping versions at query time \cite{zhu2025right,li2025tgrag}.
The same principle extends to long-context memory, where timeline-based organization links retrieved episodes to their validity windows and curbs recency bias \cite{ong2025theanine}.

\noindent\textbf{(4) Structure as a Defense to Noise.}
Flat indexing strips text of its relational and hierarchical context, leaving retrievers to operate over a bag of chunks; structural indexing restores this topology so that signal is preserved by construction.
\emph{Graph-based methods} replace arbitrary token windows with typed entity-relation networks—from GraphRAG and its variants supporting community-, subgraph-, and dual-level retrieval \cite{edge2024localtoglobal,hu2024grag,guo2024lightrag,zhu-etal-2025-knowledge}, to G-Retriever, which treats the graph itself as a first-class retrieval target \cite{he2024g-retriever}, and HippoRAG, which uses personalized PageRank over an entity graph as a long-term memory substrate \cite{hipporag}.
Because such structures only clarify when extraction is faithful, schema-aware canonicalization and retrieval-filtered construction sanitize the topology before use \cite{zhang-soh-2024-extract,zhang2025rakg}.
\emph{Hierarchical indexing} complements this by organizing evidence into recursive summary trees, enabling coarse-to-fine navigation while encapsulating local details within parent nodes for explicit source lineage \cite{sarthi2024raptor,vectifyai2025pageindex,huang2025hirag}.

\subsection{Robust Retrieval}
\label{sec:robust_retrieval}
This subsection surveys interventions that reduce retriever-level noise before context assembly.
While matching related documents from a large corpus is already a denoising process, LLM-oriented retrieval further prioritizes \emph{precision} and \emph{distractor resistance}, as the inclusion of noise can severely degrade the generation quality of LLMs.
We survey retrieval-time denoising along four dimensions: (i) query transformation, (ii) hybrid retrieval, (iii) distractor-aware retriever training, and (iv) fine-grained relevance matching.

\noindent\textbf{(1) Query Transformation.}
The mismatch between user intent and corpus representation is a starting point of retrieval noise.
Raw queries are often underspecified or ambiguous, leading retrievers to fetch topically related but factually irrelevant candidates \cite{croft2010search,ye2023enhancing}.
To bridge this semantic gap, generative expansion methods utilize LLMs to generate hypothetical documents (e.g., HyDE) or pseudo-answers, projecting the query into a clearer document embedding space \cite{gao2023precise,wang2023querydoc, yu2023generate}; or abstract the query to a higher-level concept before retrieval to evoke broader supporting evidence~\cite{zheng2024stepback}.
For complex information needs, query decomposition breaks queries into simpler sub-queries, reducing the noise associated with single-step retrieval over dense information \cite{huang2023converser,mao2023large,press2023selfask}.
Recent works further propose optimizing the rewriter via reinforcement learning (RL) or feedback loops, explicitly training rewriting to maximize end-to-end performance \cite{peng2024large,jiang2025deepretrieval,chan2024rqrag,mao-etal-2024-rafe}.

\noindent\textbf{(2) Hybrid Retrieval.}
Single retriever exhibit clear limitations: sparse methods overlook semantic nuances, while dense encoder are prone to semantic noise—retrieving topically related but factually irrelevant passages.
Hybrid retrieval addresses this by fusing dense representations with sparse lexical signals (e.g., BM25), effectively using exact matching to ground semantic associations and filter hallucinations \cite{bruch2023analysis,formal2021splade,chen2022outofdomain}.
BGE-M3 further unify this fusion into a single backbone by jointly training dense, lexical, and multi-vector heads with a hybrid scoring rule~\cite{chen2024bgem3}.
Beyond static fusion, a line work \emph{dynamically} invokes different retrieval method or LLM's parametric memoery on demands to maximize the retrieval accuracy in different scenarios~\cite{mallen-etal-2023-whennot, li2025multifield, kalra2025mor}.

\noindent\textbf{(3) Distractor-Aware Retriever Training.}
Enhancing the robustness of dense representations is critical for eliminating noise.
Several axes strengthen the retriever training against distractors. Hard negative mining force the encoder to separate genuine evidence from topical near-misses~\cite{qu2021rocketqa,wang-etal-2023-simlm,lin2023dragon}, with adversarial frameworks dynamically escalating training difficulty, exposing the retriever to ever-subtler distractor~\cite{zhang2022ar2}. Secondly, instruction tuning conditions the same embedding space on different task intents, so semantically similar but task-irrelevant passages are systematically downranked~\cite{wang2025foundation,dai2024improve,su-etal-2023-instructor}. Finally, RAG-specific objectives further distill LLM-judged utility or defects back into the retriever, aligning what is retrieved with what the generator should actually exploit~\cite{lin2024radit, tu2025rbft}.

\noindent\textbf{(4) Fine-Grained Relevance Matching.}
Compressing document into a single vector inherently cause information loss and introducing noises~\cite{weller2025theoretical}.
\emph{Late interaction} architectures like ColBERT retain token-level features for fine-grained matching, distinguishing subtle relevance from passages that are connected loosely~\cite{khattab2020colbert,santhanam2021colbertv2}.
XTR scale this paragidm by streamlining training and scoring so that token-level retrieval no longer requires gather-and-rescore, cutting cost by orders of magnitude over ColBERT~\cite{lee2023xtr}.
Methods like MUVERA directly train a multi-vector encoder to represent documents in a more fine-grained way beyond single-vector MIPS~\cite{dhulipala2024muvera}.
The same late-interaction principle has further been ported to visually-rich documents via vision-language token interaction, broadening the scope beyond single-modality RAG~\cite{faysse2025colpali}.


\subsection{Context Assembly}
\label{sec:context_assembly}
Context assembly is the first stage that take into account the interaction between retriever and generator, transforming a raw candidate set into an LLM prompt. The objective of this stage is to maximize information density under the constraints of LLM context windows. We categorize context assembly techniques into four denoising mechanisms: LLM-aware selection, prompt compression, layout optimization, and knowledge conflict resolving.

\noindent\textbf{(1) LLM-aware Selection and Pruning.}
After first-stage retrieval, reranking and pruning act as the first noise gate of context assembly, rigorously filtering irrelevant candidates before they enter the prompt.
\emph{LLM-based reranking} enhances pointwise cross-encoders \cite{nogueira2019passage} with holistic, set-level judgments:
listwise permutation generation \cite{sun2023chatgpt,ma2023zeroshot},
pairwise comparison \cite{qin2024pairwise},
setwise most-relevant selection \cite{zhuang2024setwise},
and efficiency-oriented single-token decoding \cite{reddy2024first}.
\emph{Diversity-aware subset selection} moves beyond top-$k$ truncation to
balance relevance against redundancy via MMR \cite{DBLP:conf/sigir/CarbonellG98}
or Determinantal Point Processes \cite{sarlak2026reliabilityaware}.
\emph{Fine-grained context filtering} operates at the sentence or token level
to surgically remove non-supporting content.
Speculative selection instructs the LLM itself to prune irrelevant evidence before reasoning \cite{wang2025speculative};
trained pruners \cite{wang2023filco, chirkova2025provence} distill comparable signals into lightweight classifiers,
and applies sentence-level reranking followed by LLM-based reconstruction of the surviving fragments~\cite{hwang2024dslr} .
Beyond rigid top-$k$ paradigm, adaptive strategies can dynamically size the context based on query complexity \cite{jeong2024adaptiverag}
or abstain when evidence is insufficient
\cite{penha2021calibration,meng2023query,amiraz2025distracting}.

\noindent\textbf{(2) Long Context Compression.}
To mitigate noise from verbose or irrelevant tokens, compression techniques perform \textit{soft denoising}.
\textit{Extractive compression} methods such as the LLMLingua family \cite{jiang-etal-2023-llmlingua,jiang-etal-2024-longllmlingua,pan2024llmlingua2} use small models to score token informativeness and drop low-signal tokens.
\textit{Abstractive compression} synthesizes documents into concise meta-summaries or task-specific representations \cite{xu2024recomp,li-etal-2023-compressing}.
A complementary line, \textit{soft-prompt compression}, encodes retrieved passages into continuous embedding vectors---such as Gist Tokens \cite{mu2023learning}, ICAE \cite{ge2024incontext}, and xRAG \cite{cheng2024xrag}---achieving extreme compression while preserving task performance.
All three paradigms aim to maximize information gain per token \cite{DBLP:conf/iclr/XuPSC25}, increasing the logical capacity of the context window.

\noindent\textbf{(3) Layout Optimization to Mitigate Position Bias.}
LLMs exhibit non-uniform attention, favoring the beginning and end of the context while neglecting the middle---the ``lost in the middle'' phenomenon \cite{liu2024lost}---a bias traced to causal masking and specific hidden dimensions \cite{yu2024mitigate,wang2025pine}, and amplified when relevant pieces are spaced far apart \cite{tian2025distance}.
Denoising involves \textit{topology-aware reordering} to place critical evidence at attention peaks \cite{peysakhovich2023attention}, positional calibration to flatten the U-shaped curve \cite{hsieh2024found}, and inference-time shuffling with aggregation \cite{byerly2025gold}.
These strategies dynamically structure the prompt to align with the model's inductive biases, preventing high-signal evidence from being drowned out \cite{shi2023large}.

\noindent\textbf{(4) Knowledge Conflict Resolution.}
Retrieved evidence is often internally inconsistent due to temporal drift, source disagreement, or duplicated paraphrases;
naively flattening such passages into a single prompt can lead to hallucinations or reasoning paralysis~\cite{chen2024benchmarking,liu2019generalized,xu2024knowledge}, benchmarked by~\cite{hou2024wikicontradict,su2024conflictbank}.
At assembly time, conflict-aware methods detect competing claims and route them into structured prompts that force the model to adjudicate discrepancies rather than averaging them away~\cite{xie2024adaptive,wang2025conflicting}.
Complementarily, evidence grading methods such as Chain-of-Note~\cite{yu2023chainofnote} generate passage-level reading notes (supporting vs.\ irrelevant vs.\ contradictory), while Astute RAG~\cite{wang2025astute} iteratively consolidates model's own parametric knowledge with retrieved passages according to source reliability, filtering inconsistencies before generation.
During generation, decoding-time interventions offer a second line of defense:
contrastive decoding amplifies contextual over parametric signals~\cite{shi2024trusting}, adaptive variants dynamically adjust the contrast per token to avoid over-correction of conflict~\cite{wang2025adacad}, and OpenDecoder~\cite{mo2026opendecoder} injects explicit passage quality indicators into the attention computation to steer generation toward higher-reliability passages.
\vspace{-1em}
\subsection{Retrieval Verification}
\label{sec:faithfulness_verification}
Quality and fine-grained verification are essential for improving retrieval efficacy. 
They provide critical feedback for active denoising and establish an auditable chain of evidence for human review. 
In the LLM-oriented era, retrieval evaluation extends beyond traditional hit-based metrics to encompass utility, granular traceability, and security challenges.

\noindent\textbf{(1) Evaluating Utility of Retrieval.}
Effective denoising requires evaluating whether retrieved context genuinely benefits the generation, rather than merely relevance matching in an LLM-oriented system. Traditional metrics like NDCG fail to capture the nuanced relationship between retrieval and generation quality~\cite{cuconasu2024power}.
To disentangle these effects, LLM-as-a-judge frameworks such as RAGAS \cite{es2023ragas} and RAGChecker \cite{ru2024ragchecker} separately measure components such as retrieval quality, context precision and answer faithfulness, making it possible to identify system failure modes.
Furthermore, to isolate the true information gain, utility-based metrics like SePer~\cite{daiseper} quantify the exact reduction in semantic uncertainty provided by the retrieval stage to measure retrieval utility.
Complementary benchmarks such as RAGTruth \cite{niu-etal-2024-ragtruth}, FaithEval \cite{mingfaitheval}, SafeRAG \cite{liang2025saferag}, and RGB~\cite{chen2024benchmarking} stress whether systems preserve faithfulness when evidence density degrades.

\noindent\textbf{(2) Fine-Grained Attribution of Long Answers.}
RAG systems often generate long-form answers that synthesize information across multiple retrieved passages, making it difficult for attribution. 
To localize hallucinations, recent works enforce fine-grained, span-level attribution rather than coarse document-level overlap. 
FactScore and AlignScore~\cite{min2023factscore,zha2023alignscore} assess factuality by decomposing answers into atomic claims and evaluate as aligned pairs. 
Verifiable generation methods operationalize traceability by mandating quoted evidence~\cite{menick2022gophercite, gao-etal-2023-enabling}, using critic tokens during decoding~\cite{asai2024selfrag}, or applying post-hoc verification~\cite{dhuliawala2024chainofverification}, with lightweight checkers~\cite{tang2024minicheck} further make the evaluation efficiency computationally practical.

\noindent\textbf{(3) Retrieval Safety Auditing.}
The openness of LLM-oriented IR introduces adversarial ``intentional noise'' that demands verification beyond standard quality checks. 
Retrieved documents may contain indirect prompt injections or poisoned evidence designed to hijack the generation logic, override correct evidence, or corrupt long-term agent memory \cite{liu2024promptinjection, greshake2023not,zhang2024retrievalpoisoning,zou2025poisonedrag,chen2024agentpoison}.
Against these threats, verification methods operate through complementary strategies: corroboration-based auditing such as CAR detects inconsistencies by seeking cross-source agreement under disinformation attacks~\cite{weller2024defending}; isolation-and-aggregation frameworks such as RobustRAG verify passage integrity and provide certifiable robustness guarantees against retrieval corruption~\cite{xiang2025robustrag}; and structured-query constraints expose and limit instruction-channel contamination~\cite{chen2025struq}. 
Security-focused benchmarks including SafeRAG~\cite{liang2025saferag} and RGB~\cite{chen2024benchmarking} systematically stress-test systems across diverse adversarial scenarios, transforming these verification outcomes into actionable denoising signals for upstream retrievers.

\vspace{-1em}
\subsection{Closed-Loop Training}
\label{sec:agentic_loops}
Search agents that interleave retrieval and reasoning offer capabilities beyond one-shot retrieval but are susceptible to unique vulnerabilities such as cumulative noise and error propagation. In multi-turn settings, a single irrelevant or misleading context can cascade into hallucinated reasoning, destabilizing the entire trajectory. Denoising thus shifts from static filtering to dynamic flow control. We investigate how agents mitigate noise through three mechanisms: adaptive loop orchestration, reinforcement learning, and self-evolution.
 
\noindent\textbf{Agentic loop orchestration and regulation.}
To mitigate cumulative noise, agents employ structured flow control as active noise filters.
First, query decomposition minimizes semantic drift: methods like IRCoT \cite{trivedi2023ircot} and ChainRAG \cite{zhu2025chainrag} break complex queries into verifiable sub-steps to prevent ``lost-in-retrieval'' failures, while hierarchical planners in LevelRAG \cite{zhang2025levelrag} and MindSearch \cite{chen2024mindsearch} ensure narrowly targeted retrieval to minimize distractor intrusion.
Second, adaptive scheduling acts as a dynamic gate: systems like Self-RAG \cite{asai2024selfrag}, FLARE \cite{jiang2023active}, and DRAGIN \cite{su2024dragin} assess parametric uncertainty to decide \emph{when} to retrieve, blocking unnecessary contexts that would dilute focus.
Finally, orchestration must incorporate adversarial filtering against prompt injection \cite{liu2024promptinjection} and corpus poisoning \cite{zou2025poisonedrag,shafran2025machine}, requiring interface-level sanitization such as structured queries \cite{chen2025struq}.
 
\noindent\textbf{End-to-end training for RAG.}
Rather than relying on fixed heuristics, recent works optimize denoising directly for downstream utility.
OpenRAG fine-tunes the retriever using generative feedback to filter distractors \cite{zhou2025openrag}.
Crucially, reinforcement learning (RL) aligns the entire retrieval-generation loop: by optimizing against reward signals tied to output quality, RL equips LLMs to actively seek useful evidence and ignore retrieved noise \cite{searchr1, zhang2025gvpo}.
RouteRAG employs RL to learn a router policy about when to retrieve or rely on parametric knowledge \cite{guo2025routerag}.
Similarly, Toolformer learns to execute search APIs only when they reduce perplexity, avoiding calls that introduce context noise \cite{schick2023toolformer}.
Denoising thus shifts from post-processing to a learned behavior encoded in model weights.
 
\noindent\textbf{Self-evolution and structural optimization.}
Beyond optimizing a fixed policy, agentic systems enhance denoising through \emph{continuous self-evolution}.
First, agents refine memory to prevent recurrent errors: Reflexion accumulates verbal feedback to bias trajectories away from noisy patterns \cite{shinn2023reflexion}, while MemGPT \cite{packer2023memgpt} and MemRL \cite{zhang2026memrl} implement explicit memory management---evicting irrelevant history or filtering ``similar-but-useless'' retrievals via learned utility---to maintain high signal-to-noise ratios.
Second, at the system level, AutoRAG \cite{kim2024autorag} and DSPy \cite{khattab2024dspy} treat the retrieval pipeline as a programmable surface, automatically searching for module combinations and prompt structures that maximize signal quality.
These mechanisms transition denoising from a transient tactic to a lifelong learning objective, where historical interaction and structural search drive continuous refinement of the agent's information processing pipeline.

\providecommand{\tagbox}[3]{%
  \begingroup\setlength{\fboxsep}{0.35ex}%
  \colorbox{#1}{\textcolor{#2}{\scriptsize\sffamily #3}}%
  \endgroup}
\providecommand{\stag}[1]{\tagbox{blue!12}{blue!70!black}{#1}}
\providecommand{\ftag}[1]{\tagbox{red!12}{red!75!black}{#1}}
\providecommand{\scetag}[1]{\tagbox{orange!18}{orange!85!black}{#1}}

\providecommand{\ttag}[1]{\tagbox{green!14}{green!45!black}{#1}} 
\providecommand{\jtag}[1]{\tagbox{purple!14}{purple!55!black}{#1}} 

\begin{table*}[t]
  \centering
  \small
  \setlength{\tabcolsep}{4pt}
  \renewcommand{\arraystretch}{1.18}
  \caption{
    Domain-aligned failure \& recovery cases for \emph{denoising-first} LLM-oriented IR (Sec.~4).
    \textbf{Legend:} Failure tags use \textbf{C/R/A/V/L} to denote the primary noise-entry layer (corpus/retriever/assembly/verification/loop):
    \ftag{C1} provenance contamination;
    \ftag{C2} redundancy/canonical drift;
    \ftag{C3} temporal drift/obsolescence;
    \ftag{R1} query/subquery drift;
    \ftag{R2} hard distractors;
    \ftag{A1} context dilution (lost-in-the-middle);
    \ftag{A2} low information density (verbosity/redundancy);
    \ftag{A3} evidence conflict (stale-state);
    \ftag{V1} attribution gap;
    \ftag{L1} cascading error propagation.
    Stage tags:
    \stag{CI} controlled indexing,
    \stag{RR} robust retrieval,
    \stag{CA} context assembly,
    \stag{FV} faithfulness verification,
    \stag{CL} closed-loop improvement.
  }
  \label{tab:denoise-cases}
  \begin{tabularx}{\textwidth}{@{}
    >{\raggedright\arraybackslash}p{0.125\textwidth}
    >{\raggedright\arraybackslash}p{0.18\textwidth}
    >{\raggedright\arraybackslash}p{0.182\textwidth}
    >{\raggedright\arraybackslash}X
    >{\raggedright\arraybackslash}p{0.213\textwidth}
  @{}}
    \toprule
    \textbf{Scenario} &
    \textbf{Failure Signature} &
    \textbf{Examples} &
    \textbf{Denoising Recipes} &
    \textbf{Key metrics} \\
    \midrule
    \scetag{S4.1} \textbf{Coding agents} &
    \ftag{R2} hard distractors (near-duplicate files/symbols); \ftag{L1} early mistake locks an incorrect plan; \ftag{C2} stale/deprecated code index.
    
    \textit{Symptom:} plausible patch, wrong target. &
    SWE-bench-style issue: \texttt{FAIL\_TO\_PASS} points to \texttt{src/core/parser.py}; top-$k$ retrieval ranks \texttt{src/legacy/parser.py} higher; agent patches legacy \(\rightarrow\) tests still fail. &
    \stag{RR} BM25+dense+\textit{symbol-aware} retrieval; hard-negative rerank; \stag{CI} syntax-aware repo indexing;
    \stag{CA} de-dup clones; minimal evidence (failing trace + call graph).
    \stag{FV} sandbox tests; reject mis-targeted patches.
    \stag{CL} train with distractive counterexamples. &
    \ttag{TASK} \% Resolved \cite{swebench_verified}; Pass Rate \cite{merrill2026terminal};
    \ttag{EFF} n\_turns, n\_toolcalls, n\_total\_tokens; time\_to\_last\_token.
    \ttag{TASK} ruff/mypy/bandit pass.
    \jtag{JUDGE} Agentic Rubrics \cite{raghavendra2026agentic}.
    \ttag{EFF} Cost per Task (\$);\ttag{LOC} file/function localization accuracy.\\
    \midrule
    \scetag{S4.2} \textbf{Long-term memory assistant} &
    \ftag{C2} memory store pollution
    \ftag{C3} temporal drift; 
    \ftag{A3} stale-state conflict across memories.
    
    \textit{Symptom:} response reflects outdated preference/profile. &
    \texttt{2024-03}: ``I live in Boston.'' \texttt{2024-06}: ``I moved to Seattle.''
    Later: ``recommend food near me'' \(\rightarrow\) Boston suggestions (stale memory wins). &
    \stag{CI} time-aware indexing; \stag{CI} memory consolidation;
    \stag{RR} recency-aware rerank + time-range filtering/query expansion.
    \stag{FV} state-consistency check (detect contradictions; ask/abstain). &
    \ttag{TASK} Accuracy \cite{longmemeval};
    \ttag{TASK} Answer Prediction (F1) \cite{locomo}; BLEU/ROUGE; FactScore \cite{min2023factscore};
    \ttag{RETR} Recall@K/NDCG@K.
    \jtag{JUDGE} Faithfulness \& Context Relevance \cite{es2023ragas}. \\
    \midrule
    \scetag{S4.3} \textbf{Deep research and reports} &
    \ftag{R1} subquery drift; 
    \ftag{A1} evidence dilution; 
    \ftag{V1} attribution gap.

    \textit{Symptom:} confident claims with weak/irrelevant citations. &
    Query: ``Compare PFAS regulations US vs EU (2025).''
    Subqueries drift to general health-effects; report claims ``banned globally'' with a citation to an unrelated blog snippet. &
    \stag{RR} subquery anchoring + novelty/duplicate filter.
    \stag{CA} claim-level evidence packing (``cite-by-claim'').
    \stag{FV} entailment check between each claim and cited spans.
    \stag{CL} RL-trained search policies optimize evidence density. &
    \jtag{JUDGE} RACE (Overall, Comprehensiveness, Depth, Instruction-Following, Readability).
    \ttag{CITE} FACT (Citation Accuracy,Effective Citation) \cite{deepresearchbench};
    \ttag{UTILITY} SePer \cite{daiseper}.
    \ttag{CITE} Citation Precision/Recall \cite{gao-etal-2023-enabling}. \\
    \midrule
    \scetag{S4.4} \textbf{Multimodal understanding} &
    \ftag{A2} event sparsity in long contexts; 
    \ftag{V1} weak timestamp grounding.
    
    \textit{Symptom:} misses key moment or hallucinates timestamps. &
    45-min lecture video: ``When is Theorem~2 stated?''
    Retrieval picks a 02:10 subtitle match; ground truth is 37:48 \(\rightarrow\) wrong timestamp. &
    \stag{CI} multimodal segment indexing.
    \stag{RR} dual-channel retrieval; fine-grained temporal rerank.
    \stag{CA} event-centric segment pooling/keyframe selection.
    \stag{FV} timestamp verification loop and re-query. &
    \ttag{TASK} Accuracy; F1 score;
    \ttag{RETR} Recall \cite{videomme, gao2017tall, zhou2018towards};
    \ttag{LOC} mAP (IoU=0.5/0.7) \& HIT@1 \cite{lei2021detecting};
    \ttag{TASK} MAE \& MSE \& RMSE \cite{qiu2024tfb};
    \jtag{JUDGE} MM-Relevance \cite{locomo}\\
    \bottomrule
  \end{tabularx}
\end{table*}

\section{Practices in Retrieval-Augmented Applications}
\label{sec:domain_cases}

Whereas Section\ref{sec:denoising_taxonomy} organizes denoising stage by stage, real retrieval-augmented applications fail through \emph{coupled} noise that propagates across stages rather than isolated module errors. 
We illustrate this via four representative settings—coding agents, long-term memory assistants, deep research, and multimodal understanding—and summarize their failure signatures and interventions in Table~\ref{tab:denoise-cases}.

\vspace{-1em}
\subsection{Coding Agents}
\label{sec:coding_agents}

Coding agents operate in a distinct IR environment characterized by repository-scale search spaces and rigid execution constraints. Tasks in this domain, typified by benchmarks like SWE-bench \cite{swebench,swebench_verified}, require agents to navigate massive codebases to implement precise fixes.
The central challenge is the extreme sparsity of actionable signals: a repository may contain millions of lines of code, yet the resolution to an issue often resides in a single function or a subtle cross-file dependency.

Consequently, standard retrieval is insufficient; it is plagued by high-noise distractors such as distinct functions with identical names, deprecated utilities, and complex inheritance structures that mask the true root cause.

To mitigate this, state-of-the-art workflows have evolved from flat retrieval-augmented generation into multi-stage pipelines that progressively filter noise through hierarchical localization, syntax-aware context management, and execution-based verification.

First, systems like Agentless and RepoHyper employ a coarse-to-fine filtering strategy \cite{agentless} instead of retrieving code snippets directly.
They typically start with keyword-based or vector search to identify candidate files, followed by a re-ranking or distinct selection phase where an LLM acts as a discriminator to prune irrelevant files based on the issue description. 
This hierarchical reduction transforms the retrieval problem into a decision-making process, ensuring that the limited attention budget of the downstream editing model is not diluted by structurally similar but logically unrelated code.
Furthermore, to handle the "needle-in-a-haystack" nature of cross-file dependencies, retrieval is often augmented with static analysis graphs (e.g., call graphs or import dependencies), which serve as hard constraints to guide the agent toward logically connected components rather than merely textually similarity \cite{repobench,crosscodeeval}.

Second, denoising bottleneck also lies in maximizing evidence density within the context window. 
Feeding raw source code is inefficient and noise-prone. Advanced agents address this by adopting syntax-aware representations, such as code skeletons with only class signatures and docstrings or AST-based slicing, which strip away implementation details to provide a high-level "map" of the repository \cite{sweagent,repocoder}. 
This technique, often referred to as a repository map or context compression, allows the LLM to maintain global awareness of project structure without being overwhelmed by local token noise. 
By presenting only the interface boundaries and hiding the method bodies until specifically requested, these systems artificially boost the signal-to-noise ratio, enabling the model to reason about complex architectures that would otherwise exceed its context limits.

Finally, coding agents uniquely employs execution and tests as a verification tool. 
The edit-execute loop allows agents to dynamically prune noisy trajectories using precise feedback from failing tests and stack traces. 
This signal enables systems like SWE-agent to distinguish plausible hallucinations from functional solutions, iteratively refining the retrieval context \cite{sweagent,swesearch}. 
Moreover, recent works suggest that strengthening this oracle via test augmentation or differential testing further prevents overfitting to sparse supervision \cite{utboost,patchdiff}. 
Consequently, retrieval pipeline in coding agents becomes a cyclic, verifiable loop where code structure and execution states continuously gate the flow of information.
\vspace{-1em}
\subsection{Long-term Memory Assistants}\label{sec:ltm_assistants}
Long-term memory assistants aim to maintain continuity across interaction horizons that far exceed a model's context window. 
Typically, these systems externalize history into persistent storage, yet as interactions accumulate, the raw log data inevitably degrades into a high-entropy distribution characterized by \emph{redundancy}, \emph{temporal obsolescence}, and \emph{state conflict}.
In this setting, the accumulation of history paradoxically reduces retrieval precision, creating a noisy context that leads models to hallucinate based on superseded information \cite{packer2023memgpt,locomo}.

The primary noise source in lifelong interaction is \emph{temporal drift}. 
Unlike static corpora, personal contexts are mutable; a user's preference or status recorded at $t_1$ may be directly contradicted by an event at $t_{100}$. 
Naive similarity-based retrieval often fails here, as outdated facts (``hard negatives'') share high semantic overlap with current queries, leading the LLM to hallucinate based on superseded evidence \cite{locomo,longmemeval}. 
Furthermore, indiscriminate logging creates a ``store pollution'' effect, where low-utility chitchat dilutes the retrieval pool, while adversarial inputs can introduce persistent \emph{poisoning} risks that degrade reasoning across sessions \cite{zou2025poisonedrag,minja}.

To mitigate these noise channels, denoising practices have shifted from passive indexing to active \emph{memory consolidation} and structure-aware retrieval. 
At the storage level, systems mimic human cognitive processes by abstracting raw episodic logs into synthesized semantic memories using reflection or recursive summarization to discard redundancy while preserving high-level traits \cite{generativeagents,memorybank}. 

To combat ambiguity, unstructured text is often supplemented or replaced by structured representations, such as knowledge graphs or entity-attribute pairs, which enforce stricter schemas for state tracking and enable multi-hop reasoning over connected events \cite{hipporag,licomemory}. 
At the retrieval stage, standard dense retrieval is augmented with \emph{time-awareness} via recency-weighted ranking or explicit validity filtering to ensure that the retrieved evidence reflects the current world state rather than historical artifacts \cite{memorybank}. 

Finally, during context construction, the focus lies on conflict resolution and \emph{verifiability}.
Rather than blindly packing top-$k$ chunks, advanced pipelines apply deduplication and expose provenance metadata (e.g., timestamps, source attribution), allowing the LLM to adjudicate between conflicting memories and resist poisoning attacks \cite{memorygraft,agemem}.

A shift is also happening in evaluation benchmarks, which have moved beyond simple fact recall to measuring robustness against drift and inconsistency. 
Datasets like LoCoMo \cite{locomo} and LongMemEval \cite{longmemeval} explicitly probe an agent's ability to ignore stale confounders and update beliefs in response to new evidence, underscoring that the utility of long-term memory depends less on how much is stored, and more on how effectively noise is suppressed.
\vspace{-1em}
\subsection{Deep Research}
\label{sec:deep_research}
 
Deep research systems operate over open-ended queries, heterogeneous corpora, and evolving intermediate hypotheses. Typically centered on an LLM planner and given a broad query $\mathcal{Q}$, a deep research system: (i) decomposes $\mathcal{Q}$ into a structured plan $P={g_1,\ldots,g_n}$ of subgoals, (ii) executes multi-hop retrieval over external and internal corpora to collect evidence sets $R_i$ for each $g_i$, (iii) iteratively synthesizes intermediate products while adapting the plan via dynamic replanning and meta-cognitive validation, and (iv) assembles a structured artifact with explicit citations and provenance \cite{definition,dssurvey,dssurvey2}. Unlike short-form QA, failures emerge from compounding noise across planning, retrieval, context construction, and verification, exhausting the model's attention and utilization capacity \cite{paperqa2, deepresearchbench, deepresearchslice}.
 
Noise in deep research is multi-stage and structural. At the access layer, iterative sub-query generation is prone to semantic drift: vague sub-queries retrieve topically adjacent but irrelevant literature, diluting evidence pools \cite{sage,litllm}. Even expressive LLM-based retrievers underperform due to query collapse into shallow keyword patterns, amplifying retrieval noise \cite{sage}. Domain-specific restriction and metadata augmentation partially mitigate this, indicating that denoising pressure can shift to the data layer \cite{openscholar}.
 
Context construction constitutes another bottleneck. Retrieved documents are typically long, redundant, and internally noisy; na\"{i}vely concatenating them leads to positional bias and attention collapse \cite{liu2024lost}. Multiple systems therefore introduce summarization, filtering, or re-ranking stages that transform raw documents into query-conditioned evidence snippets before insertion into the reasoning context, trading recall for evidence density and enabling agents to examine orders of magnitude more text than fits into the prompt \cite{paperqa2,searcho1}.
 
Beyond soft filtering, recent work emphasizes hard denoising mechanisms. Neuro-symbolic slicing methods formalize denoising as span selection, predicting explicit indices and discarding all other content prior to reasoning \cite{deepresearchslice}. This reframes the dominant failure mode as a retrieval--utilization gap: even when gold evidence is retrieved, it competes with overwhelming distractors. By making exclusion explicit, such methods expose utilization as a first-class bottleneck in deep research pipelines.
 
Verification and post-generation feedback introduce a critical denoising stage. Long-form reports are vulnerable to unsupported claims, citation hallucinations, and internal contradictions. Systems therefore incorporate self-review or multi-agent verification loops, auditing drafts for citation accuracy and factual consistency \cite{openscholar,literas,lira}. Benchmarks increasingly penalize spurious or missing citations, while revealing that even when retrieval succeeds, agents struggle to organize evidence into expert-like structures \cite{deepresearchbench,TaxoBench}.
 
Architecturally, deep research systems differ in where they allocate denoising effort. Planner-centric approaches reduce noise upstream by constraining subgoals and search trajectories, while draft-centric or diffusion-style systems treat the report as a noisy latent variable iteratively refined through targeted retrieval and revision \cite{TTD-DR}. Reinforcement-learning-based agents further internalize denoising by coupling outcome rewards with evidence-density or process-level signals, encouraging policies that retrieve less but better information \cite{searchr1,searchr2,deepresearcher}. Across these variants, improved performance is consistently associated with explicit mechanisms for pruning, summarizing, or verifying evidence, rather than with raw increases in retrieval breadth.
\subsection{Multimodal Understanding}
\label{sec:mm_understanding}
The transition from static text to multimodal temporal data—spanning time series, genomics, and video—introduces \textit{temporal redundancy}. Temporal streams scale linearly with duration $T$, while causal evidence $E_{causal} \subset T$ remains sparse. Thus, the retrieval challenge shifts from indexing documents to preserving causal signals within continuous, high-entropy streams.
 
In spatio-temporal modeling, naive context window expansion often degrades performance due to distribution shifts and non-stationary noise \cite{IRPA}, shifting the objective from maximizing context length to optimizing evidence density. Recent architectures treat history selection as a denoising operator: RATD \cite{RATD} and TS-RAG \cite{tsrag} replace fixed-window horizons with dynamic retrieval that fetches only segments maximizing predictive utility, while RAFT \cite{RAFT} retrieves training samples exhibiting similar temporal dynamics. To decouple short-term volatility from long-term trends, Time-MoE \cite{timemoe} routes tokens to experts specialized in specific temporal resolutions, and dual-masking strategies filter high-entropy timesteps during training to prevent overfitting to noise \cite{fu2025selective}.
 
This sparsity is more pronounced in biological and visual domains. In genomics, models like Gene42 \cite{vishniakov2025gene} scale attention to extreme lengths (e.g., 192k tokens), yet must discriminate functional regulatory motifs from vast non-coding regions. Video data represents the lower bound of signal-to-noise ratio, with relevant evidence often occupying less than 1\% of frames. The Video-MME benchmark \cite{videomme} quantifies a "long-context gap," where performance inversely correlates with duration, indicating that unpruned context acts as pollution.
 
To address this, architectures are evolving into dual-channel retrieval systems separating semantic reasoning from evidence localization. VideoRAG \cite{videorag} exemplifies a \textit{denoising-first} strategy, using Knowledge Graphs for semantic retrieval while limiting dense visual matching to relevant intervals. Complementary approaches leverage low-noise modalities (e.g., ASR transcripts) as anchors to index high-entropy visual streams. Hierarchical systems like VideoTree \cite{videotree} introduce query-adaptive pruning, filtering irrelevant branches before they reach reasoning, acting as a coarse-to-fine noise filter.
 
Finally, verifiability in multimodal generation requires explicit grounding. To prevent hallucination in temporal localization, models like VTG-LLM \cite{vtg} and TimeExpert \cite{timeexpert} decouple timestamp generation from captioning via specialized experts. By mandating timestamped citations (e.g., $t_{start}, t_{end}$), these methods transform generation into provenance-aware output, ensuring visual evidence is causally linked to the generated reasoning.

\section{Future Directions}
\label{sec:future}
We argue that LLM-oriented IR must evolve from a passive retrieval utility into an active \emph{programmable noise gate}. 
The core objective shifts from maximizing isolated recall to maximizing \emph{usable evidence density} within the model's cognitive budget.

\noindent\textbf{(1) Utility-Centric Evaluation.}
Standard ranking metrics often diverge from generation quality. 
Future benchmarks must measure \emph{causal utility}---rewarding retrieval only when it resolves reasoning gaps or corrects hallucinations---and penalize unverified attribution, enforcing a strict evidence-to-generation contract.

\noindent\textbf{(2) Proactive Index Sanitation.}
To combat the ``pollution'' of synthetic and stale content, indexing must transition to \emph{stratified governance}.
Systems should treat provenance, temporal validity, and cryptographic signatures (e.g., C2PA) as hard constraints, filtering low-credibility signals before they compete for attention.

\noindent\textbf{(3) Self-Evolving Retrieval Loops.}
Static retrievers degrade in dynamic environments. 
We envision agents that employ \emph{closed-loop feedback} to refine search policies in real-time.
By learning from downstream reasoning failures, the retriever adapts to filter hard distractors and prevent ``spiral'' errors, turning rejection signals into upstream optimization.

\noindent\textbf{(4) Optimizing Information Density in Context.}
Addressing context limitations requires moving beyond document concatenation. 
Future context assembly should operate on \emph{atomic evidence units} (e.g., specific claims, code symbols) rather than passages.
This structural compression maximizes the logical capacity of the prompt, ensuring every token contributes to verifiability.

\section{Acknowledgements}
This work was supported in part by the National Key R\&D Program of China (Grant No.2023YFF0725001), 
in part by the National Natural Science Foundation of China (Grant No. 62572417, Grant No.92370204), 
in part by the Guangdong Basic and Applied Basic Research Foundation (Grant No.2023B1515120057), 
in part by the Key-Area Special Project of Guangdong Provincial Ordinary Universities (Grant No.2024ZDZX1007).


\bibliographystyle{ACM-Reference-Format}
\bibliography{sample-base,section_3_2,section_3_4,section_4_1,section_4_2, section_4_3, section_4_4}

\appendix


\end{document}